\documentclass[referee]{raa}
\usepackage{graphicx,times}
\usepackage{natbib}
\usepackage{amssymb,amsmath}
\bibpunct{(}{)}{;}{a}{}{,}

\usepackage[a4paper=true,dvipdfm=true,pagebackref=true]{hyperref}
\hypersetup{pdftitle = The title of my PDF, pdfauthor = My name, pdfsubject= The subject, pdfkeywords = keyword1 keyword2 keyword3}
\hypersetup{colorlinks = true, linkcolor = green, anchorcolor = red, citecolor = blue, filecolor = red, pagecolor = red, urlcolor = red}

\begin{document}

   \title{EMC design for the actuators of FAST reflector
$^*$
\footnotetext{\small $*$ Supported by the National Natural Science Foundation of China(11473043).}
}

 \volnopage{ {\bf 20XX} Vol.\ {\bf X} No. {\bf XX}, 000--000}
   \setcounter{page}{1}

   \author{Hai-Yan Zhang\inst{1,2}, Ming-Chang Wu\inst{1,2}, You-Ling Yue\inst{1,2}, Heng-Qian Gan
      \inst{1,2},  Hao Hu\inst{1}, Shi-Jie Huang\inst{1}
   }

   \institute{ National Astronomical Observatories of CAS, Beijing 100012,
China; {\it hyzhang@nao.cas.cn}\\
        \and
             Key Laboratory of Radio Astronomy, CAS, Nanjing, 210034, China\\
\vs \no
   {\small Received XXXX; accepted XXXX }
}

\abstract{The active reflector is one of the three main innovations of the Five-hundred-meter Aperture Spherical radio Telescope (FAST). The deformation of such a huge spherically shaped reflector into different transient parabolic shapes is achieved by using 2225 hydraulic actuators which change the position of the 2225 nodes through the connected down tied cables. For each different tracking process of the telescope, more than 1$/$3 of these 2225 actuators must be in operation to tune the parabolic aperture accurately to meet the surface error restriction. It means that some of these actuators are inevitably located within the main beam of the receiver, and the Electromagnetic Interference (EMI) from the actuators must be mitigated to ensure the scientific output of the telescope. Based on the threshold level of interference detrimental to radio astronomy presented in ITU-R Recommendation RA.769 and EMI measurements, the shielding efficiency (SE) requirement of each actuator is set to be 80dB in the frequency range from 70MHz to 3GHz. Therefore, Electromagnetic Compatibility (EMC) was taken into account in the actuator design by measures such as  power line filters, optical fibers, shielding enclosures and other structural measures. In 2015, all the actuators had been installed at the FAST site. Till now, no apparent EMI from the actuators has been detected by the receiver, which proves the effectiveness of these EMC measures.
}

   \authorrunning{H.-Y. Zhang et al. }            
   \titlerunning{EMC design of FAST actuators}  
   \maketitle

%
\section{Introduction}           
\label{sect:intro}

   Over the past 23 years, the Five-hundred-meter Aperture Spherical radio Telescope (FAST) has been conceived, designed, optimized and constructed to provide a powerful radio tool at low frequencies for the astronomical community (Nan 2006). The active main reflector is one of the three main innovations of FAST. For each observation time-tick of 1 second, a different portion of the spherical reflector is transiently changed to a corresponding parabolic reflector with an effective aperture of 300m, focus ratio of 0.4611, a surface precision better than RMS 5mm, and optical axis pointing to the observational target with precision of 8$"$. The deformation of such a huge spherically shaped reflector into different transient parabolic shapes is achieved by using 2225 hydraulic actuators which are used to change the position of the 2225 nodes through the connected down tied cables. For each different tracking process of the telescope, about $1/3$ of these 2225 actuators must be in operation to tune the parabolic aperture accurately and meet the surface error restriction.

\begin{figure}[htb]
   \centering
   \includegraphics[width=14.0cm, angle=0]{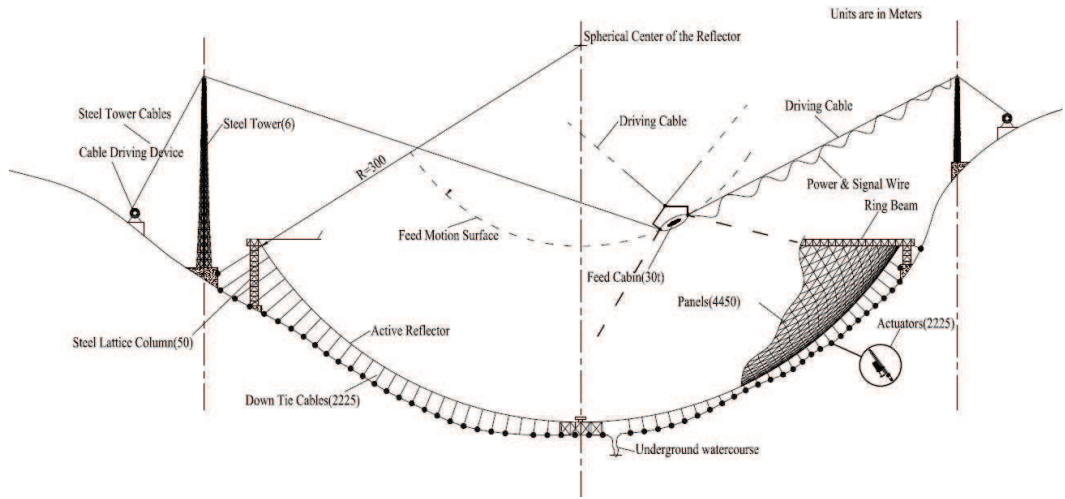}
   \caption{Schematic Diagram of FAST system. }
   \label{Fig1}
   \end{figure}

\begin{figure}[htb]
   \centering
   \includegraphics[width=7.0cm, angle=0]{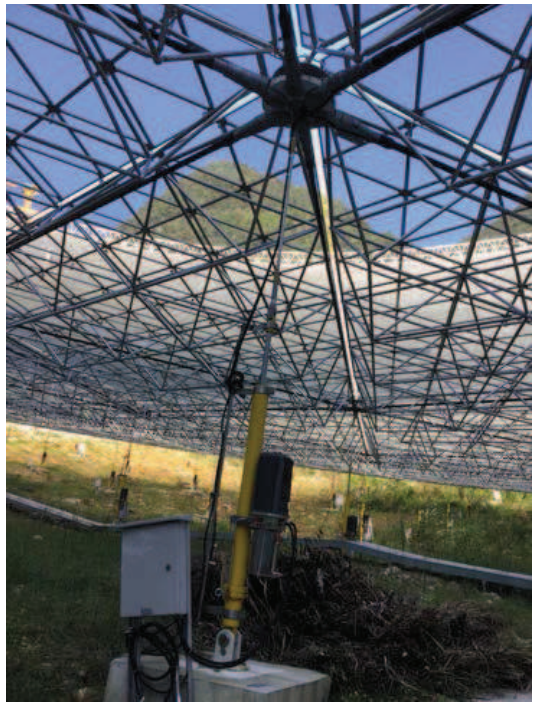}
   \caption{One actuator connected to its node plate. }
   \label{Fig2}
   \end{figure}


Due to the extreme sensitivity of FAST, the Electromagnetic Interference (EMI) issues were emphasized from the beginning of the telescope design. The most typical EMI sources for FAST are the thousands of actuators used for the deformation of the reflector, and some of these actuators are inevitably located within the main beam of the receiver, as shown in Figure.1 and 2. Therefore, it is essential for FAST to evaluate the EMI effect of the actuators and make Electromagnetic Compatibility (EMC) design for them. In the following sections, the EMC design study and testing will be described.

\section{EMC Requirement for the Actuator}
\label{sect:Req}
As the first step, the EMI of the actuator is identified and tested, which is produced by the electrical parts including the actuator controller, motor driver, power supply, integrated pressure and temperature sensor, position sensor and electromagnetic valves. In order to know the EMI properties of the actuator, measurements have been made to detect the interference from the electric parts of the actuator in the anechoic chamber according to the EMC national standard of GJB151A RE102 (GJB151A). The measurement results of the EMI from one typical actuator are presented in Figure 3. In this Figure, the EMI in the wide band has been detected and most interference is located at frequencies below 1 GHz.

The threshold level of interference detrimental to radio astronomy presented in ITU-R Recommendation RA.769 (ITU-R RA.769) has been chosen for FAST. Moreover, the nearest distance of 141m between the feed and actuator has been used to estimate the propagation loss based on the free space propagation model. In order to meet the requirements of ITU-R RA.769 in the desired frequency range, a Shielding Efficiency (SE) of about 70dB is necessary for the actuators. Since the actuators are located beneath the aluminum reflector panels, the shielding effect of the reflector had been evaluated during the design stage. Based on the theoretical analysis and practical tests by using the FAST 30m demonstrator at Miyun Station, the SE of the reflector has been estimated to be more or less than 10dB. Due to the uncertainty of this result, the shielding effect of the reflector is considered as a redundancy to ensure the protection of receiver. Moreover, considering possible long time degradation, any inconsistency of shielding measures among different actuators, and the aggregation effect, a SE requirement of 80dB is favored with a margin of about 10 to 20dB. Combining the theoretical analysis and the practical measurements, the SE requirement of the actuators is set to be 80dB in the frequency range from 70MHz to 3GHz. The typical EMI from an actuator with 80dB reduction is also presented in Figure 3.

\begin{figure}[htb]
   \centering
   \includegraphics[width=14.0cm, angle=0]{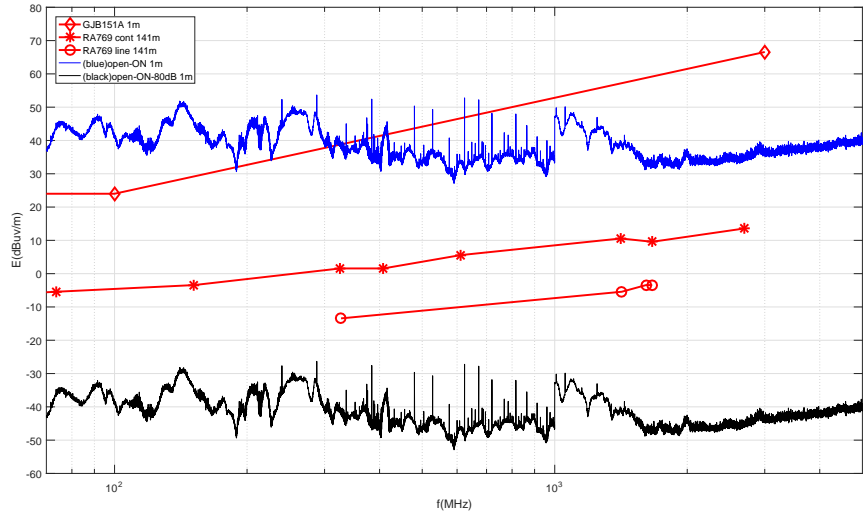}
   \caption{Typical EMI from an actuator with an open enclosure measured in the anechoic chamber based on GJB151A RE102 and with 80dB reduction. }
   \label{Fig3}
   \end{figure}

\section{EMC design of the actuator}
The overall EMC design of the actuator uses the shielding enclosure to accommodate all EMI sources. A power line filter is used to handle the AC power input. Optical fiber cables are used to transmit control and status signals. The schematic diagram of the EMC design is shown in Figure 4 and the components of the electrical part of the actuator are presented in Figure 5. The detailed design and tests for several devices are described below (Wu et al, 2015a, 2015b).

\begin{figure}[htb]
   \centering
   \includegraphics[width=14.0cm, angle=0]{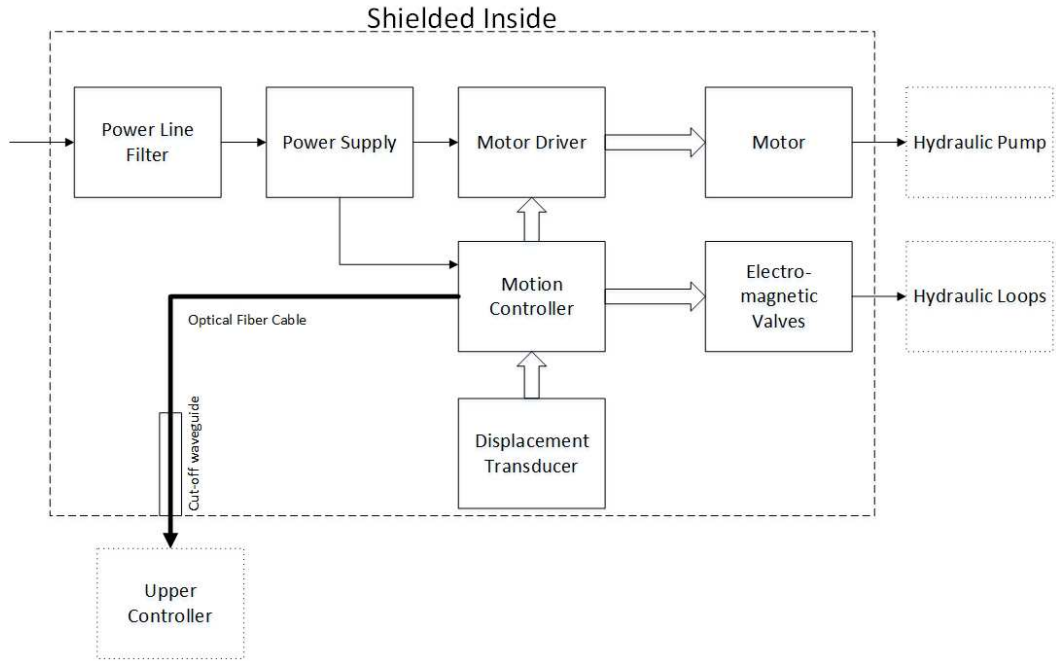}
   \caption{The schematic diagram of the EMC design. }
   \label{Fig4}
   \end{figure}

\begin{figure}[htb]
   \centering
   \includegraphics[width=14.0cm, angle=0]{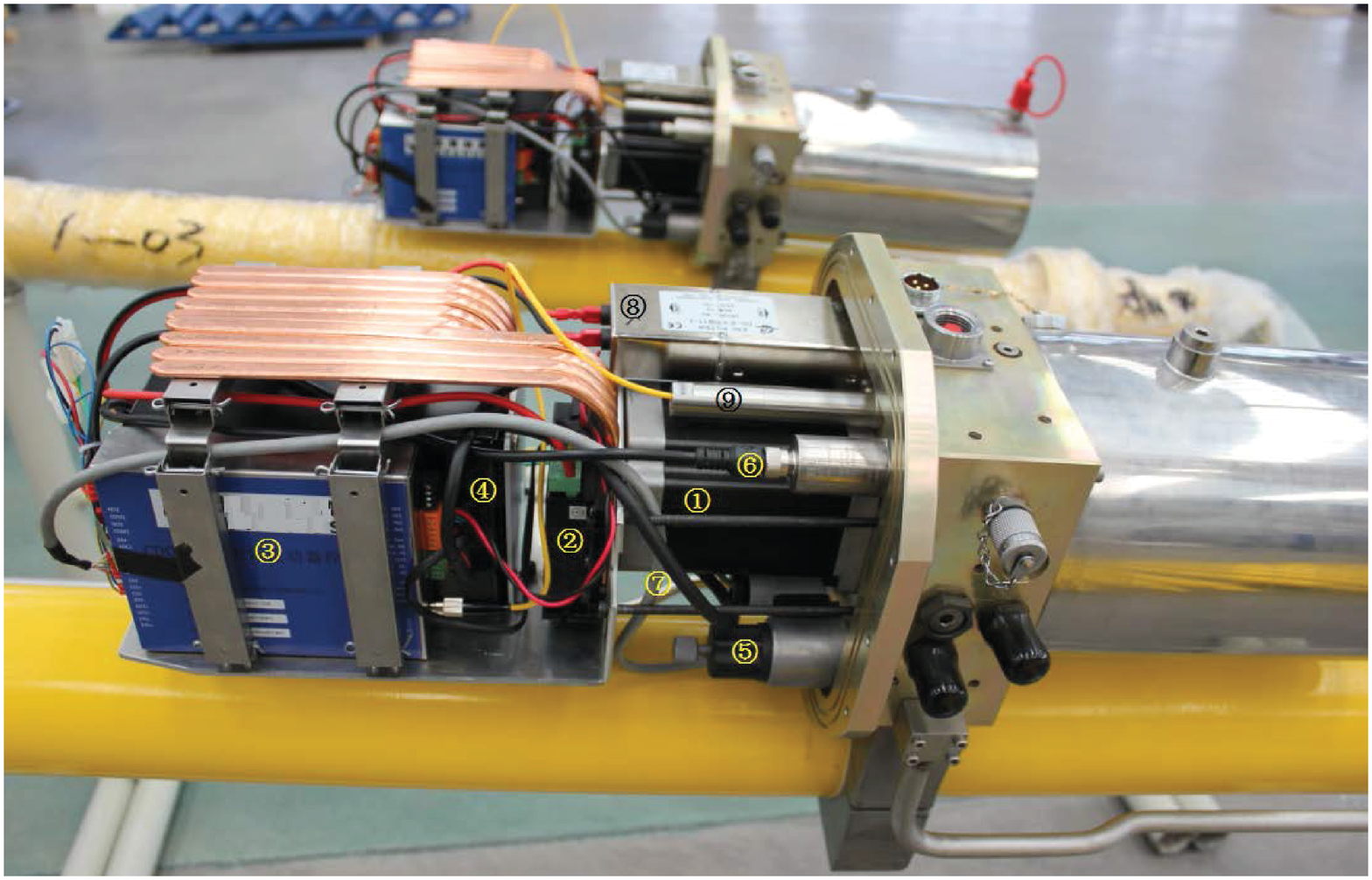}
   \caption{Electrical components inside the actuator.(1) Stepper motor; (2) Motor driver; (3) Actuator controller; (4) Power supply; (5) Electro-magnetic valve; (6) Pressure-temperature sensor; (7) Wire of the displacement sensor; (8) Power line filter; (9) Cut-off waveguide. }
   \label{Fig5}
   \end{figure}

\subsection{Power line filter}
The actuator uses AC220V 50Hz as power input. A power line filter is designed to prevent EMI transmission along the power line. The filter is designed to work at a rated voltage of 250VAC and a rated current of 6A, at the frequency of DC $\sim$ 60Hz. The insertion loss of the filter is listed in Table 1.

\begin{table}[htb]
\bc
\begin{minipage}[]{100mm}
\caption[]{The insertion loss of the filter}\end{minipage}
\setlength{\tabcolsep}{1pt}
\small
 \begin{tabular}{ccccccccccccc}
  \hline\noalign{\smallskip}
Frequency(MHz) & Insertion Loss($\geq$dB)&&&&&&&&&&\\
  \hline\noalign{\smallskip}
1 & 95 \\
5 & 100 \\
10 & 100 \\
100 & 100 \\
300$\sim$10000 & 100 \\
  \noalign{\smallskip}\hline
\end{tabular}
\ec
\end{table}

\subsection{Cut-off waveguide}
The cut-off waveguide is a stainless tube of inner diameter D=12mm and length L=115mm. The tube is installed inside the shielded enclosure, which allows the optical fiber cable to transmit control and status signals with EMI kept inside. The cutoff frequency and SE are estimated empirically to be

$f_{cutoff}=17.6/D=15(GHz)$

$SE=32L/D=32 \times 115/12=307(dB)$

where D and L are in centimeters.

\subsection{Shielding measures}
A Nickel plated graphite electroplated rubber ring has been used in the interface of the enclosure and the valve block. The material satisfies the SE of better than 80dB.

\subsection{Other EMI routes and measures}
All the cables, except the AC220V power line and the optical fiber cable, are enclosed in the corresponding enclosure. The cable of the displacement transducer is located in another enclosure at the bottom part of the cylinder. All the related EMI leak interfaces are shielded by EMC measures.

\section{Conclusion}
The actuators of the FAST reflector were manufactured after the completion of the EMC design. Before the installation at FAST site, the EMC measurements of the actuators based on the national standard of GJB151A RE102 have been made, and results show that EMI from the actuators is lower than the background of the anechoic chamber. Moreover, the SE measurements have also been carried out based on the measurement method  of Yue et al. 2015. The measurement result for one actuator is shown in Figure 6. In this figure, the actuator has met EMC requirements of 80dB in most of the band used by the receivers. In addition, each of the power line filters was tested to verify its compliance to the insertion loss requirements.

\begin{figure}[htb]
   \centering
   \includegraphics[width=14.0cm, angle=0]{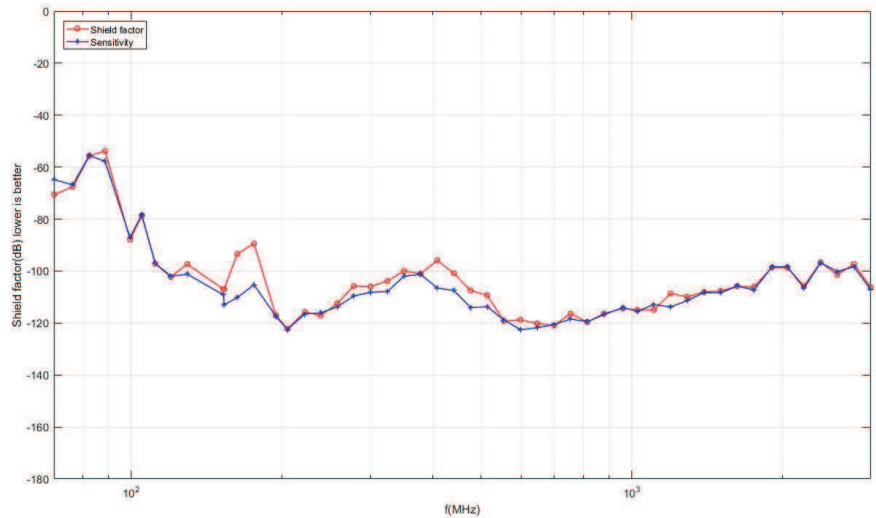}
   \caption{SE measurement result of one actuator based on the measurement method (Yue et al. 2015). }
   \label{Fig6}
   \end{figure}

For the 2225 actuators, Factory Acceptance Testing was performed. The sample rate is over 10$\%$, including more than 200 actuators.  In 2015, all the actuators had been installed at FAST site. So far, no apparent EMI from the actuators has been detected by the telescope receiver, which proves the effectiveness and success of these EMC measures.

\normalem
\begin{acknowledgements}
We are thankful for all the extraordinary efforts made by the FAST team. This work is supported by the National Natural Science Foundation of China (11473043).

\end{acknowledgements}

\label{lastpage}

\end{document}